\documentclass[letterpaper,10pt]{article} 

\usepackage{osameet3} 
\usepackage[utf8]{inputenc}
\usepackage{glossaries}
\usepackage[USenglish]{babel}

\newcommand{\SetCapsType}{normalcaps}
\usepackage{xspace}  
\usepackage[shortcuts]{extdash}
\usepackage{listofitems,pgffor}    
\usepackage{siunitx}

\makeatletter
\providecommand{\SetCapsType}{smallcaps}

\long\def\@scTrue{smallcaps}
\long\def\@scFalse{normalcaps}
\newcommand{\acroSCaps}[1]{%
    \ifx\SetCapsType\@scTrue 
        \textsc{#1}%
    \else
        \MakeUppercase{#1}%
    \fi
}
\makeatother

\usepackage{scalerel}
\makeatletter
\newcommand\scslash{%
\ifx\SetCapsType\@scTrue 
    \protect\stretchrel*{$/$}{\textsc{e}}
\else
    /
\fi
} 
\makeatother 

\newcommand{\short}[1]{%
    \glsentrytext{#1}\xspace%
}

\newcommand{\Short}[1]{%
    \Glsentrytext{#1}\xspace%
}

\newcommand{\normal}[1]{%
    \gls{#1}\xspace%
}
\newcommand{\plural}[1]{%
    \glspl{#1}\xspace%
}
\newcommand{\full}[1]{%
    \acrfull{#1}\xspace%
}
\newcommand{\Normal}[1]{%
    \Gls{#1}\xspace%
}

\newcommand{\Plural}[1]{%
    \Glspl{#1}\xspace%
}
\newcommand{\Full}[1]{%
    \Acrfull{#1}\xspace%
}

\newcommand{\nAcronym}[4][]{%
	\newacronym[#1]{#2}{#3}{#4}%
	\ifcsname#2\endcsname%
	\else%
    	\expandafter\newcommand\csname s#2\endcsname{%
    	    \ifcsname texorpdfstring\endcsname%
    	        \texorpdfstring{\short{#2}}{#2}%
    	    \else%
    	        \short{#2}%
    	    \fi%
    	    }%
    	\expandafter\newcommand\csname#2\endcsname{%
    	    \ifcsname texorpdfstring\endcsname%
    	        \texorpdfstring{\normal{#2}}{#2}%
    	    \else%
    	        \normal{#2}%
    	    \fi%
    	    }%
    	\expandafter\newcommand\csname#2s\endcsname{%
    	    \ifcsname texorpdfstring\endcsname%
    	        \texorpdfstring{\plural{#2}}{#2}%
    	    \else%
    	        \plural{#2}%
    	    \fi%
    	    }%
    	\expandafter\newcommand\csname f#2\endcsname{%
    	    \ifcsname texorpdfstring\endcsname%
    	        \texorpdfstring{\full{#2}}{#2}%
    	    \else%
    	        \full{#2}%
    	    \fi%
    	    }%
    	\expandafter\newcommand\csname su#2\endcsname{%
    	    \ifcsname texorpdfstring\endcsname%
    	        \texorpdfstring{\Short{#2}}{#2}%
    	    \else%
    	        \Short{#2}%
    	    \fi%
    	    }%
    	\expandafter\newcommand\csname u#2\endcsname{%
    	    \ifcsname texorpdfstring\endcsname%
    	        \texorpdfstring{\Normal{#2}}{#2}%
    	    \else%
    	        \Normal{#2}%
    	    \fi%
    	    }%
    	\expandafter\newcommand\csname u#2s\endcsname{%
    	    \ifcsname texorpdfstring\endcsname%
    	        \texorpdfstring{\Plural{#2}}{#2}%
    	    \else%
    	        \Plural{#2}%
    	    \fi%
    	    }%
    	\expandafter\newcommand\csname fu#2\endcsname{%
    	    \ifcsname texorpdfstring\endcsname%
    	        \texorpdfstring{\Full{#2}}{#2}%
    	    \else%
    	        \Full{#2}%
    	    \fi%
    	    }%
    \fi%
}%
 
\makeatletter
\@ifpackageloaded{babel}{%
    \newcommand{\usuk}[2]{%
        \iflanguage{USenglish}{#1}{#2}%
    }%
}{%
    \newcommand{\usuk}[2]{%
        #1%
    }%
}%
\makeatother

\DeclareRobustCommand\qamlist[1]{
    \readlist*\args{#1}%
    \acroSCaps{\args[1]\=/}%
    \ifnum\argslen = 2
        { and \acroSCaps{\args[2]}\=/}%
    \else
        \foreach \n in {2,...,\argslen}{%
            \ifnum\n = \argslen
                {, and }%
            \else 
                {, }%
            \fi
            {\acroSCaps{\args[\n]}\=/}%
        }%
    \fi
    \ifglsused{QAM}
        {}%
        {ary }%
    \ifcsname texorpdfstring\endcsname%
        \texorpdfstring{%
            {\gls{QAM}}
        }%
        {QAM}%
    \else%
        {\gls{QAM}}
    \fi
}

\newcommand\pamlist[1]{%
    \readlist*\args{#1}%
    \ifglsused{PAM}{%
        \ifcsname texorpdfstring\endcsname%
            \texorpdfstring{%
                {\gls{PAM}}%
            }%
            {PAM}%
        \else%
            {\gls{PAM}}%
        \fi%
        \acroSCaps{\=/\args[1]}%
        \ifnum\argslen = 2%
            { and \acroSCaps{\=/\args[2]}}%
        \else%
            \foreach \n in {2,...,\argslen}{%
                \ifnum\n = \argslen%
                    {, and }%
                \else
                    {, }%
                \fi%
                {\acroSCaps{\=/\args[\n]}}%
            }%
        \fi%
    }%
    {%
        \acroSCaps{\args[1]\=/}%
        \ifnum\argslen = 2
            { and \args[2]\=/}
        \else
            \foreach \n in {2,...,\argslen}{%
                \ifnum\n = \argslen
                    {, and }%
                \else
                    {, }%
                \fi
                {\acroSCaps{\args[\n]}\=/}%
            }%
        \fi
        {ary }%
        \ifcsname texorpdfstring\endcsname%
            \texorpdfstring{%
                {\gls{PAM}}
            }%
            {PAM}%
        \else%
            {\gls{PAM}}
        \fi
    }
}
 
\nAcronym{2A8PSK}{\acroSCaps{2a8psk}}{2-ary amplitude 8-ary phaseshift keying}

\nAcronym{3CCMCF}{\acroSCaps{3cc-mcf}}{3-core coupled-core multi-core fiber}

\nAcronym{4D}{\acroSCaps{4d}}{four-dimensional}
\nAcronym{4D64PRS}{\acroSCaps{4d-64prs}}{\usuk{four-dimensional 64-ary polarization-ring-switching}{four-dimensional 64-ary polarisation-ring-switching}}

\nAcronym{5B4D2A8PSK}{\acroSCaps{5b4d-2a8psk}}{5-bit four-dimensional two-amplitude 8-ary phase-shift keying}

\nAcronym{6B4D2A8PSK}{\acroSCaps{6b4d-2a8psk}}{6-bit four-dimensional two-amplitude 8-ary phase-shift keying}

\nAcronym{8D}{\acroSCaps{8d}}{eight-dimensional}\nAcronym{8D2048PRS}{\acroSCaps{8d-2048prs}}{eight-dimensional 2048-ary polarization-ring-switching}
\nAcronym{8D2048PRST1}{\acroSCaps{8d-2048prs-t1}}{eight-dimensional 2048-ary polarization-ring-switching type 1}
\nAcronym{8D2048PRST2}{\acroSCaps{8d-2048prs-t2}}{eight-dimensional 2048-ary polarization-ring-switching type 2}

\nAcronym{ABC}{\acroSCaps{abc}}{automatic bias controller}
\nAcronym{AC}{\acroSCaps{ac}}{alternating current}
\nAcronym{ADC}{\acroSCaps{adc}}{analog-to-digital converter}
\nAcronym{AIR}{\acroSCaps{air}}{achievable information rate}
\nAcronym{AOM}{\acroSCaps{aom}}{acoustic optical modulator}
\nAcronym{API}{\acroSCaps{api}}{application programming interface}
\nAcronym{AR}{\acroSCaps{ar}}{achievable rate}
\nAcronym{ASE}{\acroSCaps{ase}}{amplified spontaneous emission}
\nAcronym{ASK}{\acroSCaps{ask}}{amplitude-shift keying}
\nAcronym{ASIC}{\acroSCaps{asic}}{application-specific integrated circuit}
\nAcronym{ARRWG}{\acroSCaps{a}rr\acroSCaps{wg}}{arrayed-waveguide grating}
\nAcronym{AWG}{\acroSCaps{awg}}{arbitrary-waveform generator}
\nAcronym{AWGN}{\acroSCaps{awgn}}{additive white Gaussian noise}

\nAcronym{BCH}{\acroSCaps{bch}}{Bose-Chaudhuri-Hocquenghem}
\nAcronym{BER}{\acroSCaps{ber}}{bit error rate}
\nAcronym{BICM}{\acroSCaps{bicm}}{bit-interleaved coded modulation}
\nAcronym{BMD}{\acroSCaps{bmd}}{bit-metric decoding}
\nAcronym{BPD}{\acroSCaps{bpd}}{balanced photo-diode}
\nAcronym{BPF}{\acroSCaps{bpf}}{bandpass filter}
\nAcronym{BPS}{\acroSCaps{bps}}{blind phase search}
\nAcronym{BPSK}{\acroSCaps{bpsk}}{binary phase-shift keying}
\nAcronym{BRGC}{\acroSCaps{brgc}}{binary reflected Gray code}
\nAcronym{BTB}{\acroSCaps{btb}}{back-to-back}

\nAcronym{CAGR}{\acroSCaps{cagr}}{compound annual growth rate}
\nAcronym{CCDM}{\acroSCaps{ccdm}}{constant composition distribution matching}
\nAcronym{CCF}{\acroSCaps{ccf}}{\usuk{coupled-core fiber}{coupled-core fibre}}
\nAcronym{CD}{\acroSCaps{cd}}{chromatic dispersion}
\nAcronym{CIR}{\acroSCaps{cir}}{channel impulse response}
\nAcronym{CMA}{\acroSCaps{cma}}{constant modulus algorithm}
\nAcronym{CMUX}{\acroSCaps{cmux}}{core multiplexer}
\nAcronym{ChUT}{\acroSCaps{chut}}{channel under test}
\nAcronym{CUT}{\acroSCaps{cut}}{channel under test}
\nAcronym{CPE}{\acroSCaps{cpe}}{carrier phase estimation}
\nAcronym{CPU}{\acroSCaps{cpu}}{central processing unit}
\nAcronym{CSPR}{\acroSCaps{cspr}}{carrier-to-signal power ratio}
\nAcronym{CUDA}{\acroSCaps{cuda}}{compute unified device architecture}
\nAcronym{CW}{\acroSCaps{cw}}{continuous wave}
\nAcronym{CCD}{\acroSCaps{ccd}}{charge-coupled device}

\nAcronym{DA}{\acroSCaps{da}}{driver amplifier}
\nAcronym{DAC}{\acroSCaps{dac}}{digital-to-analog converter}
\nAcronym{DC}{\acroSCaps{dc}}{direct current}
\nAcronym{DCF}{\acroSCaps{dcf}}{\usuk{dispersion compensated fiber}{dispersion compensated fibre}}
\nAcronym{DCI}{\acroSCaps{dci}}{data center interconnect}
\nAcronym{DDLMS}{\acroSCaps{dd-lms}}{decision-directed least mean square}
\nAcronym{DEMUX}{\acroSCaps{demux}}{de-multiplexer}
\nAcronym{DFA}{\acroSCaps{dfa}}{\usuk{doped fiber amplifier}{doped fibre amplifier}}
\nAcronym{DFB}{\acroSCaps{dfb}}{distributed feedback}
\nAcronym{DGD}{\acroSCaps{dgd}}{differential group delay}
\nAcronym{DH}{\acroSCaps{dh}}{digital holography}
\nAcronym{DM}{\acroSCaps{dm}}{distribution matcher}
\nAcronym{DMA}{\acroSCaps{dma}}{direct memory access}
\nAcronym{DMD}{\acroSCaps{dmd}}{differential mode delay}
\nAcronym{DMG}{\acroSCaps{dmg}}{differential modal gain}
\nAcronym{DMGD}{\acroSCaps{dmgd}}{differential mode group delay}
\nAcronym{DML}{\acroSCaps{dml}}{directly-modulated laser}
\nAcronym{DP}{\acroSCaps{dp}}{\usuk{dual-polarization}{dual-polarisation}}
\nAcronym{DPC}{\acroSCaps{dpc}}{digital pre-compensation}
\nAcronym{DPE}{\acroSCaps{dpe}}{digital pre-emphasis}
\nAcronym{DPIQ}{\acroSCaps{dp-iqm}}{\usuk{dual-polarization \acroSCaps{iq}-modulator}{dual-polarisation \acroSCaps{iq}-modulator}}
\nAcronym{DPLL}{\acroSCaps{dpll}}{digital phase-locked loop}
\nAcronym{DQPSK}{\acroSCaps{dqpsk}}{differential quaternary phase-shift-keying}
\nAcronym{DRA}{\acroSCaps{dra}}{distributed Raman amplifier}
\nAcronym{DRE}{\acroSCaps{dre}}{digital resolution enhancer}
\nAcronym{DSF}{\acroSCaps{dsf}}{\usuk{dispersion-shifted fiber}{dispersion-shifted fibre}}
\nAcronym{DSO}{\acroSCaps{dso}}{digital sampling oscilloscope}
\nAcronym{DSP}{\acroSCaps{dsp}}{digital signal processing}
\nAcronym{DUT}{\acroSCaps{dut}}{device-under-test}
\nAcronym{DWDM}{\acroSCaps{dwdm}}{dense wavelength-division multiplexing}

\nAcronym{EAM}{\acroSCaps{eam}}{electro-absorption modulator}
\nAcronym{ECL}{\acroSCaps{ecl}}{external cavity laser}
\nAcronym{ED}{\acroSCaps{ed}}{Eucledian distance}
\nAcronym{EDF}{\acroSCaps{edf}}{\usuk{erbium-doped fiber}{erbium-doped fibre}}
\nAcronym{EDFA}{\acroSCaps{edfa}}{\usuk{erbium-doped fiber amplifier}{erbium-doped fibre amplifier}}
\nAcronym{ENOB}{\acroSCaps{enob}}{effective number of bits}
\nAcronym{ESS}{\acroSCaps{ess}}{enumerative sphere shaping}

\nAcronym{FD}{\acroSCaps{fd}}{frequency domain}
\nAcronym{FDE}{\acroSCaps{fde}}{\usuk{frequency domain equalizer}{frequency domain equaliser}}
\nAcronym{FEC}{\acroSCaps{fec}}{forward error correction}
\nAcronym{FFT}{\acroSCaps{fft}}{fast Fourier transform}
\nAcronym{FIR}{\acroSCaps{fir}}{finite impulse response}
\nAcronym{FMEDF}{\acroSCaps{fmedf}}{\usuk{few-mode erbium-doped fiber}{few-mode erbium-doped fibre}}
\nAcronym{FMEDFA}{\acroSCaps{fm-edfa}}{\usuk{few-mode erbium-doped fiber amplifier}{few-mode erbium-doped fibre amplifier}}
\nAcronym{FMF}{\acroSCaps{fmf}}{\usuk{few-mode fiber}{few-mode fibre}}
\nAcronym[plural=FM-MCF, firstplural=\usuk{few-mode multi-core fibers}{few-mode multi-core fibres}]{FM-MCF}{\acroSCaps{fm-mcf}}{\usuk{few-mode multi-core fiber}{few-mode multi-core fibre}}
\nAcronym{FPGA}{\acroSCaps{fpga}}{field-programmable gate array}
\nAcronym{FWM}{\acroSCaps{fwm}}{four-wave mixing}
\nAcronym{FSO}{\acroSCaps{fso}}{free-space optics}

\nAcronym{GD}{\acroSCaps{gd}}{group delay}
\nAcronym{GI}{\acroSCaps{gi}}{graded-index}
\nAcronym{GFF}{\acroSCaps{gff}}{gain flattening filter}
\nAcronym{GIMMF}{\acroSCaps{gi-mmf}}{\usuk{graded-index multi-mode fiber}{graded-index multi-mode fibre}}
\nAcronym{GMI}{\acroSCaps{gmi}}{\usuk{generalized mutual information}{generalised mutual information}}
\nAcronym{GNSE}{\acroSCaps{gnse}}{generalized nonlinear Schr\"{o}dinger equation}
\nAcronym{GPU}{\acroSCaps{gpu}}{graphics processing unit}
\nAcronym{GS}{\acroSCaps{gs}}{geometric shaping}
\nAcronym{GV}{\acroSCaps{gv}}{group-velocity}
\nAcronym{GVD}{\acroSCaps{gvd}}{group-velocity dispersion}
\nAcronym{GPIO}{\acroSCaps{gpio}}{general purpose input output}
\nAcronym{GUI}{\acroSCaps{gui}}{graphical user interface}

\nAcronym{HDFEC}{\acroSCaps{hd-fec}}{hard decision forward error correction}
\nAcronym{HV}{\acroSCaps{hv}}{Hufnagel-Valley}
\nAcronym{HAP}{\acroSCaps{hap}}{Hufnagel-Andrew-Phillips}

\nAcronym[plural=IL, firstplural=insertion losses (\acroSCaps{il})]{IL}{\acroSCaps{il}}{insertion loss}
\nAcronym{IFFT}{\acroSCaps{ifft}}{inverse fast Fourier transform}
\nAcronym{IMDD}{\acroSCaps{im}\scslash \acroSCaps{dd}}{intensity-modulation direct-detection}
\nAcronym{IQM}{\acroSCaps{iqm}}{in-phase and quadrature modulator}
\nAcronym{ISI}{\acroSCaps{isi}}{inter-symbol interference}

\nAcronym{JGN}{\acroSCaps{jgn}}{Japan Gigabit Network}

\nAcronym{KK}{\acroSCaps{kk}}{Kramers-Kronig}

\nAcronym{LCOS}{\acroSCaps{LCoS}}{liquid crystal on silicon}
\nAcronym{LDPC}{\acroSCaps{ldpc}}{low-density parity-check}
\nAcronym{LEAF}{\acroSCaps{leaf}}{\usuk{large effective area fiber}{large effective area fibre}}
\nAcronym{LFSR}{\acroSCaps{lfsr}}{linear-feedback shift register}
\nAcronym{LMS}{\acroSCaps{lms}}{least means squares}
\nAcronym{LLR}{\acroSCaps{llr}}{log-likelihood ratio}
\nAcronym{LO}{\acroSCaps{lo}}{local oscillator}
\nAcronym{LP}{\acroSCaps{lp}}{\usuk{linearly polarized}{linearly polarised}}
\nAcronym{LSPS}{\acroSCaps{lsps}}{\usuk{loop-synchronized polarization scrambler}{loop-synchronised polarisation scrambler}}
\nAcronym{LUT}{\acroSCaps{lut}}{lookup table}

\nAcronym{MB}{\acroSCaps{mb}}{Maxwell-Bolzmann}
\nAcronym{MCF}{\acroSCaps{mcf}}{\usuk{multi-core fiber}{multi-core fibre}}
\nAcronym{MDG}{\acroSCaps{mdg}}{mode dependent gain}
\nAcronym[plural=MDL, firstplural=mode-dependent losses (\acroSCaps{mdl})]{MDL}{\acroSCaps{mdl}}{mode-dependent loss}
\nAcronym{MDM}{\acroSCaps{mdm}}{mode-division multiplexing}
\nAcronym{MEMS}{\acroSCaps{mems}}{micro-electro-mechanical systems}
\nAcronym{MF}{\acroSCaps{mf}}{matched filter}
\nAcronym{MI}{\acroSCaps{mi}}{mutual information}
\nAcronym{MIMO}{\acroSCaps{mimo}}{multiple-input multiple-output}
\nAcronym{ML}{\acroSCaps{ml}}{machine learning}
\nAcronym{MMA}{\acroSCaps{mma}}{multi-modulus algorithm}
\nAcronym{MMEDF}{\acroSCaps{mmedf}}{\usuk{multi-mode erbium-doped fiber}{multi-mode erbium-doped fibre}}
\nAcronym{MMEDFA}{\acroSCaps{mmedfa}}{\usuk{multi-mode erbium-doped fiber amplifier}{multi-mode erbium-doped fibre amplifier}}
\nAcronym{MMF}{\acroSCaps{mmf}}{\usuk{multi-mode fiber}{multi-mode fibre}}
\nAcronym{MMSE}{\acroSCaps{mmse}}{minimum mean squared error}
\nAcronym{MP}{\acroSCaps{mp}}{minimum phase}
\nAcronym{MPLC}{\acroSCaps{mplc}}{multi-plane light converter}
\nAcronym{MSE}{\acroSCaps{mse}}{mean squared error}
\nAcronym{MUX}{\acroSCaps{mux}}{multiplexer}
\nAcronym{MZM}{\acroSCaps{mzm}}{Mach-Zehnder modulator}
\nAcronym{MZI}{\acroSCaps{mzi}}{Mach-Zehnder interferometer}

\nAcronym{NA}{\acroSCaps{na}}{numerical aperture}
\nAcronym{NF}{\acroSCaps{nf}}{noise figure}
\nAcronym{NGMI}{\acroSCaps{ngmi}}{\usuk{normalized generalized mutual information}{normalised generalised mutual information}}
\nAcronym{NLSE}{\acroSCaps{nlse}}{nonlinear Schr\"{o}ding equation}
\nAcronym{NN}{\acroSCaps{nn}}{neural network}
\nAcronym{NICT}{\acroSCaps{nict}}{National Institute of Information and Communications Technology}
\nAcronym{NIR}{\acroSCaps{nir}}{near-infrared}

\nAcronym{OBTB}{\acroSCaps{obtb}}{optical back-to-back}
\nAcronym{ODE}{\acroSCaps{ode}}{ordinary differential equation}
\nAcronym{ODL}{\acroSCaps{odl}}{optical delay line}
\nAcronym{OFC}{\acroSCaps{ofc}}{Optical Fiber Communications Conference}
\nAcronym{OFDR}{\acroSCaps{ofdr}}{optical frequency-domain reflectometer} 
\nAcronym{OFDM}{\acroSCaps{ofdm}}{orthogonal frequency division multiplexing}
\nAcronym{OH}{\acroSCaps{oh}}{overhead}
\nAcronym{OMFT}{\acroSCaps{omft}}{optical multi-format transmitter}
\nAcronym{OOK}{\acroSCaps{ook}}{on-off keying}
\nAcronym{OPLL}{\acroSCaps{opll}}{optical phase-locked loop}
\nAcronym{OSA}{\acroSCaps{osa}}{\usuk{optical spectrum analyzer}{optical spectrum analyser}}
\nAcronym{OSNR}{\acroSCaps{osnr}}{optical signal-to-noise ratio}
\nAcronym{OTDR}{\acroSCaps{otdr}}{optical time-domain reflectometer}
\nAcronym{OTF}{\acroSCaps{otf}}{optical tunable filter}
\nAcronym{OVNA}{\acroSCaps{ovna}}{\usuk{optical vector network analyzer}{optical vector network analyser}}
\nAcronym{OTG}{\acroSCaps{otg}}{optical turbulence generator}

\nAcronym{PAM}{\acroSCaps{pam}}{pulse-amplitude modulation}
\nAcronym{PAS}{\acroSCaps{pas}}{probabilistic amplitude shaping}
\nAcronym{PAPR}{\acroSCaps{papr}}{peak-to-average power ratio}
\nAcronym{PBC}{\acroSCaps{pbc}}{\usuk{polarization beam combiner}{polarisation beam combiner}}
\nAcronym{PBS}{\acroSCaps{pbs}}{polarization beam splitter}
\nAcronym{PCVD}{\acroSCaps{pcvd}}{plasma chemical vapor depostion}
\nAcronym{PCG}{\acroSCaps{pcg64}}{64-bit permuted congruential generator}
\nAcronym{PD}{\acroSCaps{pd}}{photodiode}
\nAcronym{PDL}{\acroSCaps{pdl}}{polarization-dependent loss}
\nAcronym{PDM}{\acroSCaps{pdm}}{\usuk{polarization-division multiplexing}{polarisation-division multiplexing}}
\nAcronym{PL}{\acroSCaps{pl}}{photonic lantern}
\nAcronym{PMBPSK}{\acroSCaps{pm-bpsk}}{polarization-multiplexed binary phase-shift-keying}
\nAcronym{PMQPSK}{\acroSCaps{pm-qpsk}}{polarization-multiplexed quaternary phase-shift-keying}
\nAcronym{PM8QAM}{\acroSCaps{pm-8qam}}{polarization-multiplexed 8-ary quadrature amplitude modulation}
\nAcronym{PMD}{\acroSCaps{pmd}}{polarization mode dispersion}
\nAcronym{PMF}{\acroSCaps{pmf}}{\usuk{polarization-maintaining fiber}{polarisation-maintaining fiber}}
\nAcronym{PNOB}{\acroSCaps{pnob}}{physical number of bits}
\nAcronym{PON}{\acroSCaps{pon}}{passive-optical network}
\nAcronym{PRBS}{\acroSCaps{prbs}}{pseudorandom bit sequence}
\nAcronym{PS}{\acroSCaps{ps}}{probabilistic shaping}
\nAcronym{PSK}{\acroSCaps{psk}}{phase-shift keying}
\nAcronym{PSP}{\acroSCaps{psp}}{\usuk{principal states of polarization}{principal states of polarisation}}
\nAcronym{PID}{\acroSCaps{pid}}{proportional–integral–derivative}
\nAcronym{PV}{\acroSCaps{pv}}{process value}

\nAcronym{QAM}{\acroSCaps{qam}}{quadrature amplitude modulation}
\nAcronym{QPSK}{\acroSCaps{qpsk}}{quaternary phase-shift-keying}

\nAcronym{RAM}{\acroSCaps{ram}}{random-access memory}
\nAcronym{RC}{\acroSCaps{rc}}{raised cosine}
\nAcronym{RF}{\acroSCaps{rf}}{radio frequency}
\nAcronym{RLS}{\acroSCaps{rls}}{recursive least squares}
\nAcronym{RRC}{\acroSCaps{rrc}}{root-raised-cosine}
\nAcronym{ROADM}{\acroSCaps{roadm}}{reconfigurable optical add-drop multiplexer}
\nAcronym{ROI}{\acroSCaps{roi}}{region of interest}

\nAcronym{SA}{\acroSCaps{sa}}{simulated annealing}
\nAcronym{SamPerSym}{\acroSCaps{sps}}{samples per symbol}
\nAcronym{SBS}{\acroSCaps{sbs}}{stimulated Brillouin scattering}
\nAcronym{SDFEC}{\acroSCaps{sd-fec}}{soft decision forward error correction}
\nAcronym{SDM}{\acroSCaps{sdm}}{space-division multiplexing}
\nAcronym{SE}{\acroSCaps{se}}{spectral efficiency}
\nAcronym{SER}{\acroSCaps{ser}}{symbol error rate}
\nAcronym{SI}{\acroSCaps{si}}{step index}
\nAcronym{SLM}{\acroSCaps{slm}}{spatial light modulator}
\nAcronym{SMF}{\acroSCaps{smf}}{\usuk{single-mode fiber}{single-mode fibre}}
\nAcronym{SNR}{\acroSCaps{snr}}{signal-to-noise ratio}
\nAcronym{SOA}{\acroSCaps{soa}}{semiconductor optical amplifier}
\nAcronym[firstplural=\usuk{states of polarization (\acroSCaps{sop})}{states of polarisation (\acroSCaps{sop})}]{SOP}{\acroSCaps{sop}}{\usuk{state of polarization}{state of polarisation}}
\nAcronym{SPM}{\acroSCaps{spm}}{self-phase modulation}
\nAcronym{SPS}{\acroSCaps{sps}}{samples per symbol}
\nAcronym{SRS}{\acroSCaps{srs}}{stimulated Raman scattering}
\nAcronym{SSBI}{\acroSCaps{ssbi}}{signal-signal beat interference}
\nAcronym{SSFM}{\acroSCaps{ssfm}}{split-step Fourier method}
\nAcronym{SSMF}{\acroSCaps{ssmf}}{\usuk{standard single-mode fiber}{standard single-mode fibre}}
\nAcronym{STL}{\acroSCaps{stl}}{swept tunable laser}
\nAcronym{SVD}{\acroSCaps{svd}}{singular value decomposition}
\nAcronym{SWI}{\acroSCaps{swi}}{swept wavelength interferometry}
\nAcronym{SP}{\acroSCaps{sp}}{setpoint}

\nAcronym{TD}{\acroSCaps{td}}{time domain}
\nAcronym{TDE}{\acroSCaps{tde}}{\usuk{time domain equalizer}{time domain equaliser}}
\nAcronym{TDM}{\acroSCaps{tdm}}{time-domain multiplexing}
\nAcronym{TDMSDM}{\acroSCaps{tdm-sdm}}{time-domain multiplexed space-division multiplexing}
\nAcronym{TE}{\acroSCaps{te}}{transverse electric}
\nAcronym{TIA}{\acroSCaps{tia}}{trans-impedance amplifier}
\nAcronym{TM}{\acroSCaps{TM}}{transverse magnetic}
\nAcronym{TH4D}{\acroSCaps{th-4d}}{time domain hybrid four-dimensional}
\nAcronym{TH4D2A8PSK}{\acroSCaps{th-4d-2a8psk}}{time domain hybrid four-dimensional two-amplitude eight-phase shift keying}

\nAcronym{VCSEL}{\acroSCaps{vcsel}}{vertical-cavity surface emitting laser}
\nAcronym{VHDL}{\acroSCaps{vhdl}}{\acroSCaps{Vhsic} Hardware Description Language}
\nAcronym{VOA}{\acroSCaps{voa}}{variable optical attenuator}

\nAcronym{WDM}{\acroSCaps{wdm}}{wavelength-division multiplexing}
\nAcronym{WGA}{\acroSCaps{wga}}{weakly guiding approximation}
\nAcronym{WGN}{\acroSCaps{wgn}}{white Gaussian noise}
\nAcronym[firstplural=wavelength selective switches (\acroSCaps{wss}s)]{WSS}{\acroSCaps{wss}}{wavelength selective switch}

\nAcronym{XPM}{\acroSCaps{xpm}}{cross-phase modulation}
\nAcronym{XT}{\acroSCaps{xt}}{cross-talk}

\nAcronym{3DWG}{\acroSCaps{3dwg}}{3D-waveguide} 

\usepackage{xcolor}
\usepackage{float}
\AtEndPreamble{\DeclareSymbolFont{operators}{OT1}{\rmdefault}{m}{n}}
\usepackage{siunitx}
\usepackage{lipsum}

\usepackage{silence}
\usepackage[subrefformat=parens]{subcaption}

\usepackage{amsmath,amssymb}
\usepackage[colorlinks=true,bookmarks=false,citecolor=blue,urlcolor=blue]{hyperref} 
\usepackage[capitalise]{cleveref}

\begin{document}

\title{Optical Field Characterization using Off-axis Digital Holography}

\author{Sjoerd~van~der~Heide\textsuperscript{(1)},
        Bram~van~Esch\textsuperscript{(1)},
        Menno~van~den~Hout\textsuperscript{(1)},
        Thomas Bradley\textsuperscript{(1)},
        Amado~M.~Velazquez-Benitez\textsuperscript{(1),(4)}
        Nicolas K. Fontaine\textsuperscript{(2)},
        Roland Ryf\textsuperscript{(2)},
        Haoshuo Chen\textsuperscript{(2)},
        Mikael Mazur\textsuperscript{(2)},
        Jose~Enrique~Antonio-L\'opez\textsuperscript{(3)},
        Juan~Carlos Alvarado-Zacarias\textsuperscript{(3)}, 
        Rodrigo~Amezcua-Correa\textsuperscript{(3)},
        and Chigo~Okonkwo\textsuperscript{(1)}
}
\address{
 \textsuperscript{(1)} High-Capacity Optical Transmission Laboratory, Eindhoven University of Technology, 5600 MB, Eindhoven\\
 \textsuperscript{(2)} Nokia Bell Labs, 600 Mountain Ave, New Providence, NJ 07974, USA\\
 \textsuperscript{(3)} CREOL, The College of Optics and Photonics, University of Central Florida, Orlando, 32816, USA\\
 \textsuperscript{(4)} Instituto de Ciencias Aplicadas y Tecnología, Universidad Nacional Autónoma de México, Circuito Exterior S/N, Ciudad Universitaria, 04510, Mexico City.
 }
\email{s.p.v.d.heide@tue.nl}


\begin{abstract*}
Angular resolved digital holography is presented as a technique for real-time characterization of the full optical field (amplitude and phase) of space-division multiplexing components and fibers, here a 6-mode photonic-lantern is characterized.
\end{abstract*}

\section{Overview}
Recent experimental demonstrations have shown the potential of \SDM to greatly increase fiber transmission data rates by multiplexing spatial paths \cite{Puttnam2021}. \uMDM, a subset of \SDM, uses modes of \FMFs or \MMFs as independent spatial paths. The optical field transmitted through these fibers is much more complex than in conventional \SMFs, making its characterization challenging. Current characterization methods include analyzing equalizer taps \cite{Hout2020a} or using an \OVNA \cite{Rommel2017}, both require demultiplexing into single-mode domain. Therefore, these methods limit characterization to the modal basis enabled by the (de)multiplexer, for example a \PL \cite{Velazquez-Benitez2018}, which is often not well matched to the modal basis of the \DUT. Furthermore, the measurements of the \DUT are distorted by the demultiplexer. Off-axis \gls{DH} is able to directly measure the optical field, enabling characterization without (de)multiplexer. In many recent publications where multi-mode devices are discussed \cite{Heide2020a,Heide2020c,Alvarado-Zacarias2020,Mazur2019,Mazur2021,Mounaix2020,Mounaix2019,Fontaine2019a,Fontaine2019}, \gls{DH} is used to characterize the device.

In this demo, we use \gls{DH} to measure optical fields in a \FMF to characterize a 6-mode \PL. The angular polarization multiplexing \gls{DH} setup introduced in \cite{Heide2020a} is improved through inserting the reference beams via mirrors, removing the need for beam splitters in the optical path. The improved setup shows excellent temporal stability and low wavelength dependence.

\begin{figure*}[!b]
\vspace{-5mm}
\centering
\includegraphics[width=\textwidth]{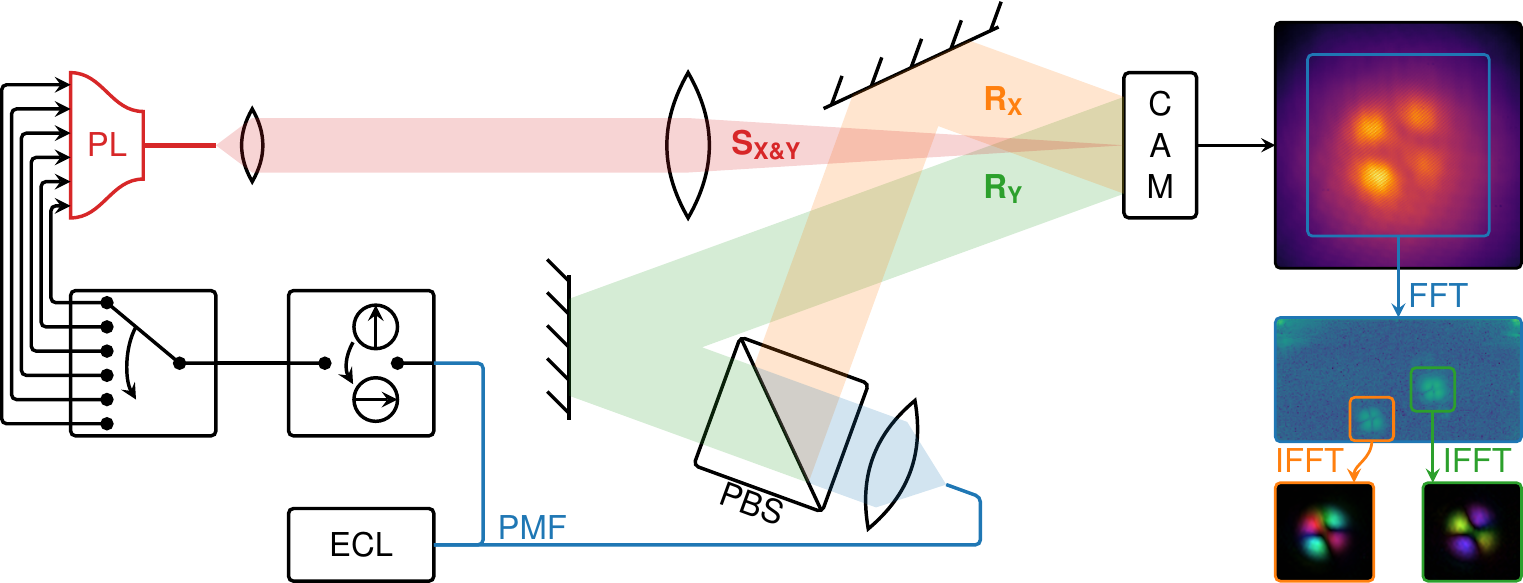}
\caption{Experimental setup. PL: photonic lantern, S\textsubscript{X\&Y}: dual-polarization signal to be characterized, R\textsubscript{X} and R\textsubscript{Y}: reference beam for x- and y-polarization, FFT: fast Fourier transform, ECL: external cavity laser.}
\label{fig:expsetup}
\vspace{-5mm}
\end{figure*}

\newpage
\section{Innovation}
\SDM components and fibers have an inherently greater level of complexity in their optical characterization than \SMFs. With increasing research levels around \MDM it is becoming necessary to have real-time advanced characterization tools which are sophisticated enough to characterize the full optical field exiting from single components but additionally from full systems of concatenated \SDM components and fibers. Angular resolved DH allows characterization of the full complex optical field, while giving insight into \MDL and \XT at a component and system level. Here, we detailed the developments in our angular resolved DH system which allow real-time, stable measurement of \FMFs and \FMF devices with an high accuracy and repeatability. The innovations can be summarized as follows, decreased temporal sensitivity to external perturbations and hence increased reliability and repeatability through precise control of the reference polarization and phase fronts. This is combined with custom \DSP for real-time processing and phase front compensation.

\cref{fig:expsetup} shows the experimental off-axis \gls{DH} setup where light exiting the facet of the \PL few-mode output fiber is imaged on to a CCD camera using a 4f optical setup with lenses with focal distances of \SIlist{4.5;750}{mm}, providing 167x magnification. Off-axis \gls{DH} requires a flat-phase coherent reference beam with a slight angle (\SI{\sim 5}{\degree}) with respect to the signal beam to construct a fringe pattern on the camera, after which digital analysis reveals not only the amplitude but also the phase of the signal beam. Only co-polarized signals can interfere, therefore, two reference beams with different angles and orthogonal polarizations are provided to support polarization-diverse measurements. The right-hand side of \cref{fig:expsetup} shows a camera frame and a simplified explanation of the \DSP chain. The camera captures the intensity of the optical field, containing the interference between signal (S\textsubscript{X\&Y}) and x-, and y-polarization references (R\textsubscript{X} and R\textsubscript{Y}) and can be described by

\begin{equation*}
    |S_{X\&Y} + R_X + R_Y|^2 = S_{X}R_X^* + S_{Y}R_Y^* + |S_{X\&Y}|^2 + |R_X|^2 + |R_Y|^2
\end{equation*}

\begin{figure*}
\centering
\includegraphics[width=\textwidth]{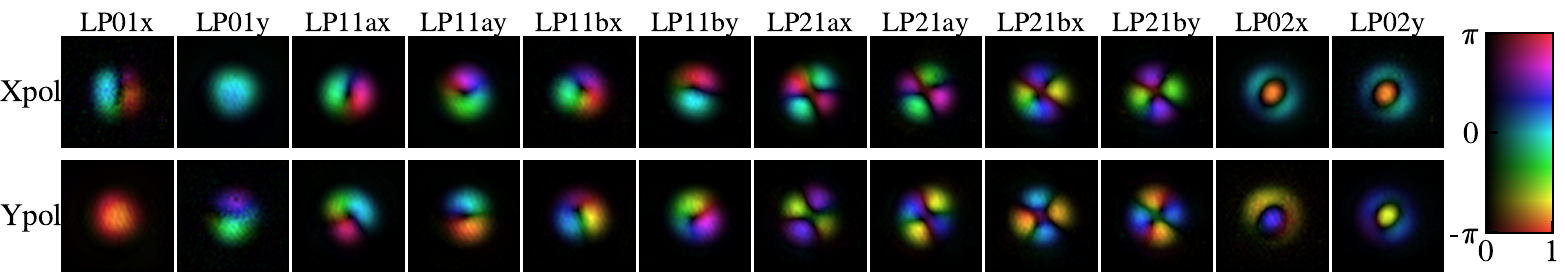}
\caption{Measured x- and y-polarization optical fields for every input port and polarization of the photonic lantern.}
\label{fig:modepictures}
\end{figure*}

A real-valued \FFT converts from the spatial domain to the angular domain and reveals the interference pattern for both polarizations. The signal-signal and reference-reference beating terms appear at or near DC. The two signal-reference beating terms are at an offset from DC, controlled by the angle between them. Both interference patterns are individually cropped and converted back to spatial domain using an \IFFT and now contain phase information. The measurement provides both amplitude and phase information for both polarizations, but only for a single input state of the \PL. Two optical switches are used to repeat the measurement for every input port and input polarization combination of the \PL, the results of which are shown in \cref{fig:modepictures}. Note that if results from different ports are jointly analyzed, the system is assumed stationary during the measurement, hence, any pertubations during the measurement will introduce errors in the analysis. To improve stability and strengthen the stationary assumption, the switches and camera operate very fast, switching every \SI{3}{ms} and completing the measurement of a 6-mode \PL in \SI{36}{ms}.

\begin{figure*}
    \centering
    \begin{subfigure}{0.49\textwidth}
        \includegraphics{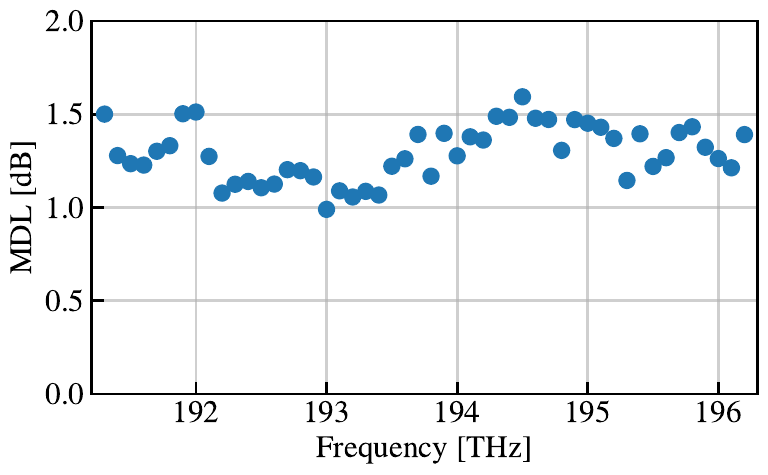}
        \caption{Stable \sMDL measured across the C-band.}
        \label{fig:MDLresults}
    \end{subfigure}%
    \hfill%
    \begin{subfigure}{0.49\textwidth}
        \includegraphics{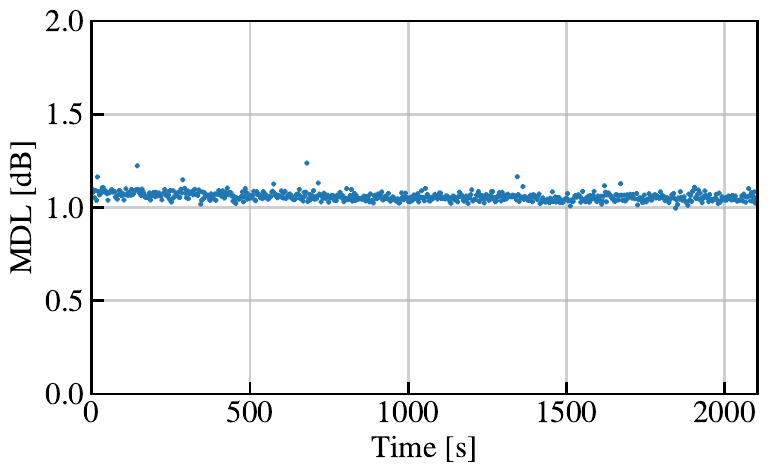}
        \caption{MDL versus time, showing stable results over half an hour.}
        \label{fig:MDLtimeresults}
    \end{subfigure}%
    \caption{Experimental \sMDL results of a 6-mode photonic lantern obtained using an off-axis \gls{DH} setup.}
    \label{fig:resultstop2}
\end{figure*}

The results of the measurements, shown in \cref{fig:modepictures}, can be used for further processing. The measured optical signals can be digitally demultiplexed into any desired modal basis by calculating overlap integrals with that modal basis. Each overlap integral calculates one entry of a transfer matrix between the measured \PL and the desired modal basis. \uSVD of this transfer matrix is used to calculate \MDL. \cref{fig:MDLresults} shows the \MDL as a function of optical frequency, showing good results for the C-band. \cref{fig:MDLtimeresults} shows the \MDL versus time at \SI{193.4}{THz} or \SI{1550.116}{nm}, showing stable results for over half an hour.

At this point, we do not know whether the \MDL variation of \cref{fig:MDLresults} is due to the wavelength dependence of the \PL or the measurement apparatus, since we cannot distinguish between them. However, compared to the setup introduced in \cite{Heide2020a}, beam splitters have been removed and results are more stable across the C-band. Also, the removal of beam splitters removed many reflections in the setup, making alignment easier. The setup in \cite{Heide2020a} contained two collimators, one for each reference beam, which were physically rotated to make references orthogonal. This turned out to make the setup unstable. First, any movement of the fibers to the collimators introduced a phase difference between reference polarizations, increasing \MDL, requiring the fibers to be taped down rigorously even though \PMF was used. Second, any polarization rotation in the fibers leading to the reference arms was propagated to the camera, making the system unstable, again even though \PMF was used. Now, a single collimator is used in conjunction with a \PBS, decoupling any polarization rotations before the \PBS from the references incident on the camera. These changes have made the setup much more stable over time.

\section{OFC Relevance}
\Gls{DH} has become an invaluable tool in our laboratory. We use it not only to characterize devices, but also to help assembly of free-space optical setups and fiber devices. For example, in \cite{Alvarado-Zacarias2020}, \gls{DH} was used to assemble a multi-mode \EDFA. The device required fiber tapers which introduce lots of scattering and unrecoverable \MDL if errors occur during fabrication. Splices, especially between different fiber types, can introduce \MDL due to modal mismatch, misalignment, and fabrication errors. \Gls{DH} enables measuring \MDL at every stage of assembly, spotting many errors which would have otherwise not been discovered until the entire \EDFA was assembled. Therefore, we believe \gls{DH} is a great addition to any laboratory working on fiber devices or free-space optical setups.

\section{Demo content \& implementation}
An \FMF coupling scenario will be showcased where \gls{DH} helps a researcher to align two \FMFs, showing \MDL and transfer matrices updated in quasi real-time. The demonstration will be held remotely, with a live video link showing a computer screen and a live camera view of the laboratory. Attendees will be able to talk to the demonstrator and ask questions. Posters with schematics and photographs of the setup will be physically posted next to the computer screens at OFC. A video will be made available for on-demand attendees.

\vspace{1mm}
\noindent
\emph{\footnotesize{\noindent Partial funding is from  the Dutch NWO Gravitation Program on Research Center for Integrated Nanophotonics (Grant Number 024.002.033), from the KPN-TU/e Smart Two program and from the Dutch NWO Visitor's program (Grant number 040.11.743).}}

\bibliographystyle{osajnl}
\bibliography{ref.bib}

\end{document}